# Multi-agent architecture for supply chain management

Roy Daniel, Anciaux Didier, Monteiro Thibaud, Ouzizi Latifa

MACSI – INRIA, LGIPM – AGIP
Université de Metz, île du Saulcy, F-57012 Metz CEDEX 01 – FRANCE
Email: {roy, anciaux, monteiro, ouzizi}@agip.sciences.univ-metz.fr

**Abstract:**

The purpose of this paper is to propose a new approach for the supply chain management. This approach is based on the virtual enterprise paradigm and the used of multi-agent concept.

Each entity (like enterprise) is autonomous and must perform local and global goals in relation with its environment. The base component of our approach is a Virtual Enterprise Node (VEN).

The supply chain is viewed as a set of tiers (corresponding to the levels of production), in which each partner of the supply chain (VEN) is in relation with several customers and suppliers. Each VEN belongs to one tier.

The main customer gives global objectives (quantity, cost and delay) to the supply chain. The Mediator Agent (MA) is in charge to manage the supply chain in order to respect those objectives as global level. Those objectives are taking over to Negotiator Agent at the tier level (NAT). These two agents are only active if a perturbation occurs; otherwise information flows are only exchange between VENs.

This architecture allows supply chains management which is completely transparent seen from simple enterprise of the supply chain. The used of Multi-Agent System (MAS) allows physical distribution of the decisional system. Moreover, the hierarchical organizational structure with a decentralized control guaranties, in the same time, the autonomy of each entity and the whole flexibility.

## 1. Introduction

In today's increasing global and competitive marketplace, it is imperative that enterprises work together to achieve the expected goals in terms of minimizing the delay of deliveries, the holding costs and the transportation costs. New forms of organizations have emerged, the so-called extended enterprises and virtual enterprises, in which partners must demonstrate strong co-ordination and commitment capabilities to achieve the desired goals. A Virtual Enterprise could be a single enterprise or a regrouping of similar companies (i.e. similar goods).

Today, in a supply chain, manufacturers no longer produce complete products in isolated facilities. They operate as nodes (i.e. single or virtual enterprise) in a network of suppliers, customers, warehouse and other specialized service functions (Davidow, & Malone, 1995).

Due to the high complexity of a whole supply chain, a centralized decisional system seems not be able to manage easily all the necessary information and actions. More over, the centralized philosophy is strongly opposed to the decisional autonomy of the supply-chain components (firms). This is why, we propose a more distributed approach in order to adhere to nodes autonomy and to facilitate the management.

Supply chain management needs to integrate two decision levels: planning and control. In planning a supply chain, coherent planning of all actors is needed. This integration not only applies to the material flows from raw material suppliers to finished product delivery, but also to the financial flows and information flows from the market (i.e. the anonymous consumers) back to the supply-chain partners. This planning function lies at the tactical level of the supply chain. Control function has a shorter run decision and a smaller focus than planning. Its objectives are restricted on one single or virtual enterprise. It lies at the operational level.

The purpose of this paper is to propose a new approach for the supply chain management. This approach is based on the virtual enterprise paradigm and the used of multi-agent concept.

The virtual enterprise is defined as a regrouping of nodes (or entities) which are linked together with information and material flows. Of course, each node could be itself a virtual or simple enterprise.

First, the supply chain architecture is defined. We detail our approach in the second part, giving attention to the supply chain three main levels: actors, tiers and global. Then, we focus our approach on the elementary actor level. And finally, we present some conclusions and further research.

## 2. Supply chain context

### 2.1. Architecture definition

The supply chain considered in our work can be summarized as follows (Figure 1):

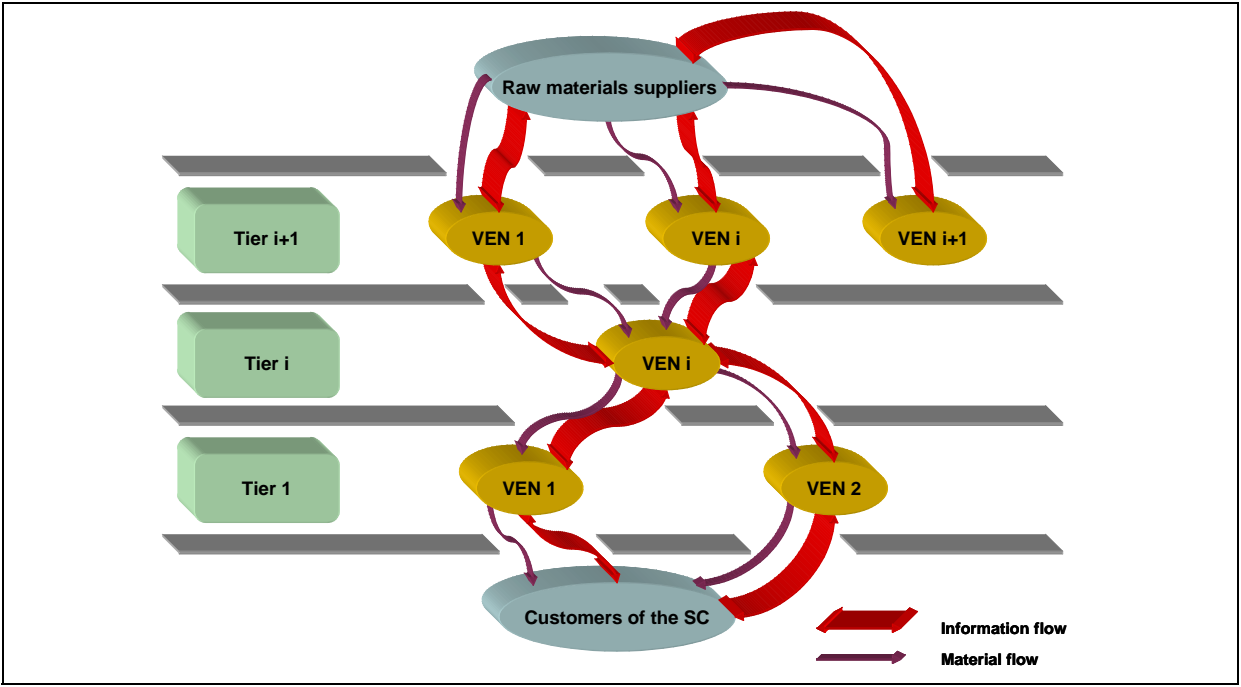

*Figure 1. Supply chain architecture*

The supply chain is viewed as a set of tiers (corresponding to the nomenclature levels), in which each partner, called a virtual enterprise node (VEN), is in relation with customers and suppliers on the adjacent tiers. We assume that each VEN is only in relationship with its adjacent VENs (no loop between the VENs allowed). So, each VEN belongs to one tier.

The VEN is the base component of this architecture. As mention before, this VEN could be a single enterprise or a regrouping of similar companies (i.e. similar goods). In that case, each company could transfer a part of its production to others.

The concept of virtual enterprise (VE) was introduced with an aim of widening the concept of the extended enterprise towards a concept of less centralized organization. Contrary to the extended enterprise which is arranged around a decision center, the VE characterizes an independent consortium which links their ressources for grow their reactivity regarding the unpredictable environment (Hardwick, & Bolton, 1997).

It is assumed that any component can be provided to each VEN by only one supplier VEN without the possibility of changing it by another supplier, except in the case of a long-term disagreement.

To generate a better productivity, these companies need the coordination of the actions which are distributed among autonomous partners (Altersohn, 1992; Rota, 1998; Kjenstad, 1998). Recent research shows a growing interest in studying cooperation relationships among the multiple actors of an industrial architecture (Axelrod, 1992; Rapoport, 1987; Ferrarini, 2001; Monteiro, & Ladet, 2001b). Cooperation can take various forms. It can be defined as a collaboration between partners, each having equivalent decisional capacity and acting together towards a common objective. One example of collaboration is the co-design in the automotive (Womack, Jones, & Roos, 1992) or aeronautical sectors. Cooperation can also be defined as the coordination and synchronization of operations carried out by independent actors (Malone, & Crowston, 1994; Monteiro, & Ladet, 2001a). Each partner has a limited decision power that corresponds to its action field (Camalot, Esquirol, Huguet, & Erschler, 1997; Camalot, 2000; Huguet, 1994).

## *2.2. Cost definition and cost negotiation*

The problem of VENs is to coordinate decentralized actions and to establish a coherent planning in real time (planning taking into account the variation of forecasting and capacity of VENs). So, in the same time, VENs may assure local and global benefits within the supply chain. Locally, we are interested in optimizing the behavior of each VEN; the objective of each VEN is to minimize purchasing and production costs and also to ensure a positive benefit (Anciaux, Ouzizi, & Portmann, 2003).

The global benefit of the supply chain is:

$$\sum_{all\ VENs} selling - \sum_{all\ VENs} costs \geq 0 \qquad (1)$$

## 3. Management architecture definition

In order to benefit, in the same time, from the agility of the distributed approach and from the coordination of the centralized one, we choose to mix these two approaches in a specific architecture.

### *3.1. Multi agent paradigm*

The intrinsic distribution of the supply chain cannot be managed by only one and single data-processing application. Indeed, the exchanges of information and the behaviors specific to operations of the supply chain members are so complex that they ask for CPU time which can only be shared. Accordingly, it seems to us that only a Multi-Agent architecture can meet this need (Ferber, 1995; Patriti, Schäfer, Ramos, Charpentier, Martin, & Veron, 1997).

The supply chain is modeled as a multi-agent system, agents use cooperative negotiation to establish coherent decisions. To limit the negotiation process in terms of iterations, a Negotiator Agent for each Tier *j* of the virtual enterprise ($NAT_j$) and a Mediator Agent (MA) for the whole virtual enterprise will be used.

The architecture of the system is given by Figure 2.

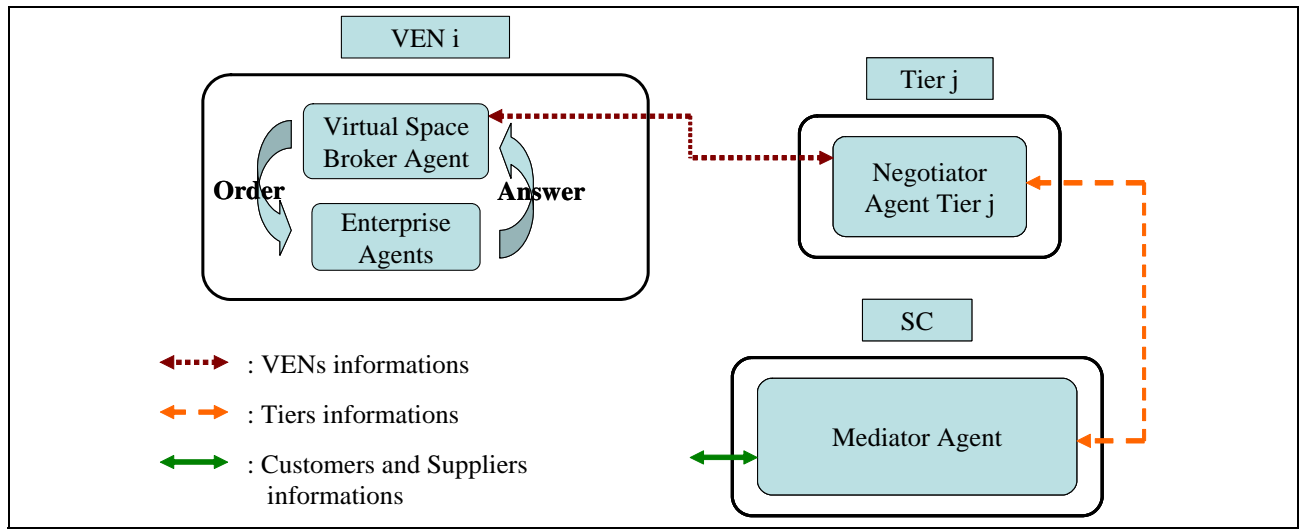

*Figure 2. Agent organization*

## *3.2. VEN level*

VEN is the elementary component of our architecture. VEN is composed by a single or many companies. In that case, those ones gather around a project, either to face competition, or to answer a request largely higher than the individual partners' capacities. These companies commonly exchange among themselves a part or all their orders. This allows each to solve the reoccurring problem of internal production overload. Even if, seen outside, only one company is in direct relation with the customer; it is in fact several companies which answer together the initial offer.

### 3.2.1. VEN organization

With the reception of a new order from a client, a member of a VEN, qualified Principal Supplier (PS), evaluates in-house his capacity of answer. If this one cannot ensure the entire induced load, it contacts its partners to absorb the overload. Our approach consists in using for this purpose a Virtual Space Broker Agent (VSBA) to manage the overload distribution among the VEN.

A client's order must contain the following information:
- Client and Principal Supplier identifiers (Ct, PS)
- Product i references ($P_i$)
- Order characteristics: quality (K), maximum cost ($C_{max}$), quantity (Q) and due date (D)

A split delivery is possible. In this case, a minimum quantity Qmin has to be delivered at least for D, the rest having to be delivered at the latest to Dmax. Figure 3 illustrated this research space (Monteiro, & Roy, 2003).

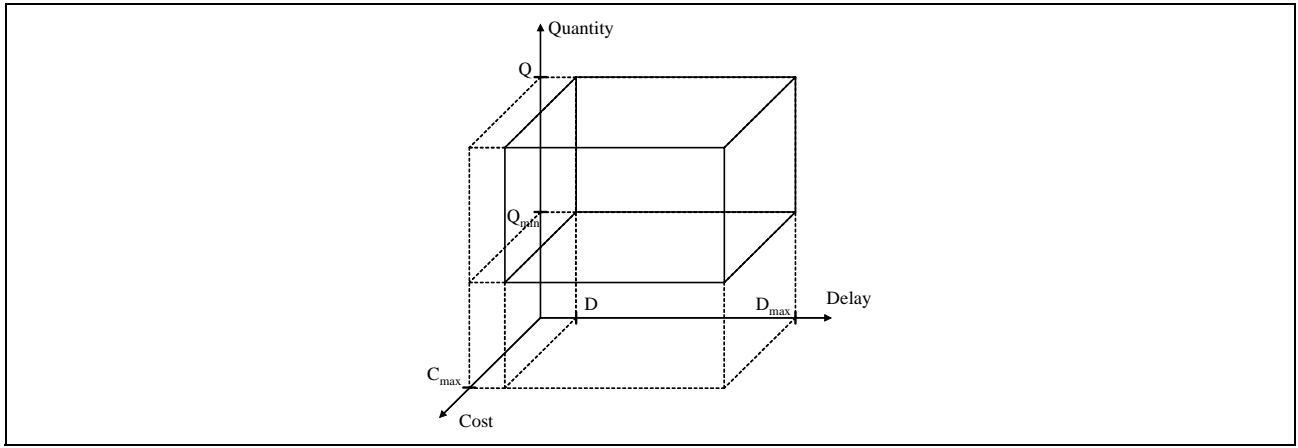

*Figure 3. Research space*

The choice is based on an algorithm for demand estimation (Monteiro, Ladet, & Bouchriha, 2003). The goal is to decide, using knowledge of the systems current state, if the company (PS) can accept or reject a new demand from its client (Ct). This decision is based on the evaluation of the load on a production center. To determine rapidly in what conditions the company is able to manufacture the new order, a comparison is made between the added load induced by this new manufacturing demand and the idle (unused production capacity) of each planning period.

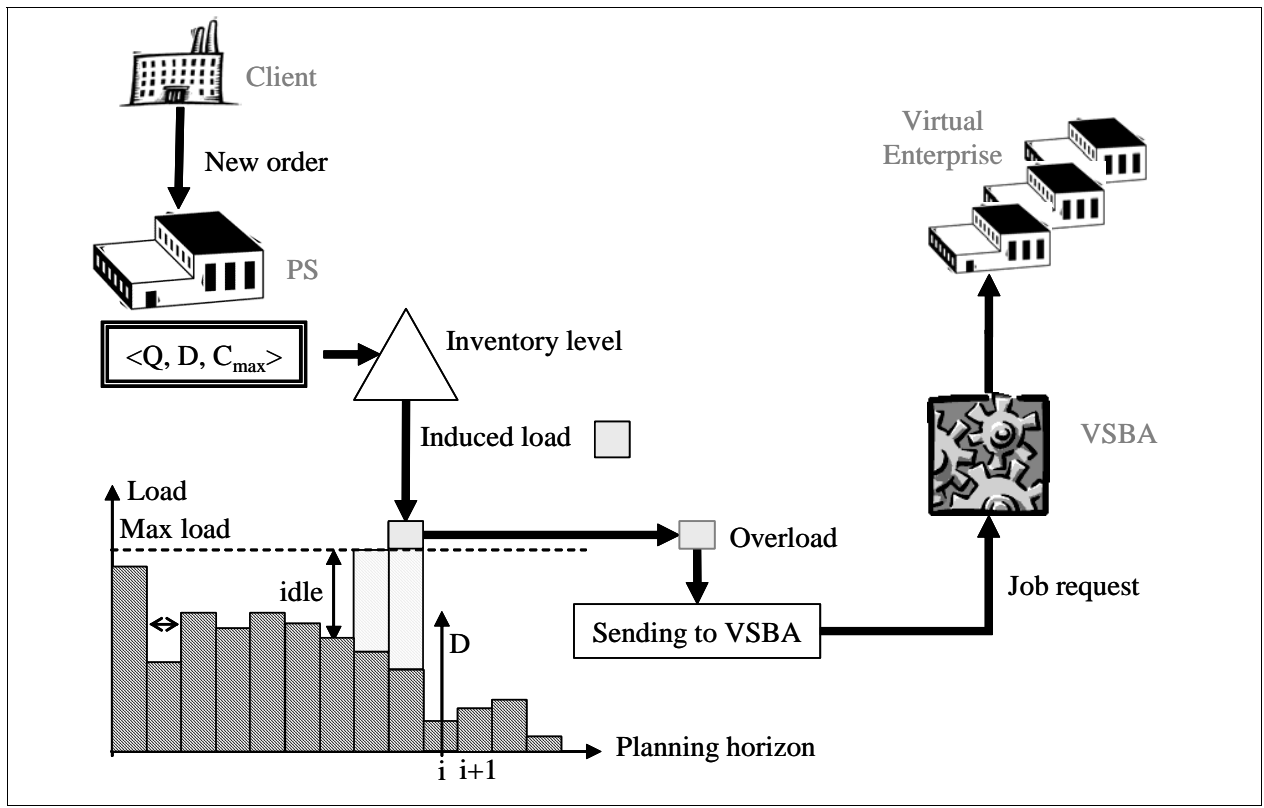

*Figure 4. Load of a production center and new demand estimation process*

This analysis is done by focusing on the bottleneck activity[1] of the internal production system.

---

[1] We call bottleneck activity, activity which most restrains the load in the internal production process.

### 3.2.2. The Virtual Space Broker Agent (VSBA)

The VSBA receives from one of its partners (PS for this command) an initial demand (Job request) which contains the following limits for:
- Quantity (Q', Q'$_{min}$),
- Delay (D, D$_{max}$),
- Cost (C'$_{max}$)

Those limits are fixed by the PS according to the part that the PS makes itself.

Then, the VSBA reflects the request on the various partners by forwarding request for production without overtime for the date D. So, VSBA send requests with following form : <Q', D, HS=0>. Each partner i could reply to this request by sending its proposal in terms of quantity and cost (qi, ci). That information is used for VSBA's decision. Following the algorithm described in the Figure 5, VSBA is able to answer to the PS and share the job between partners.

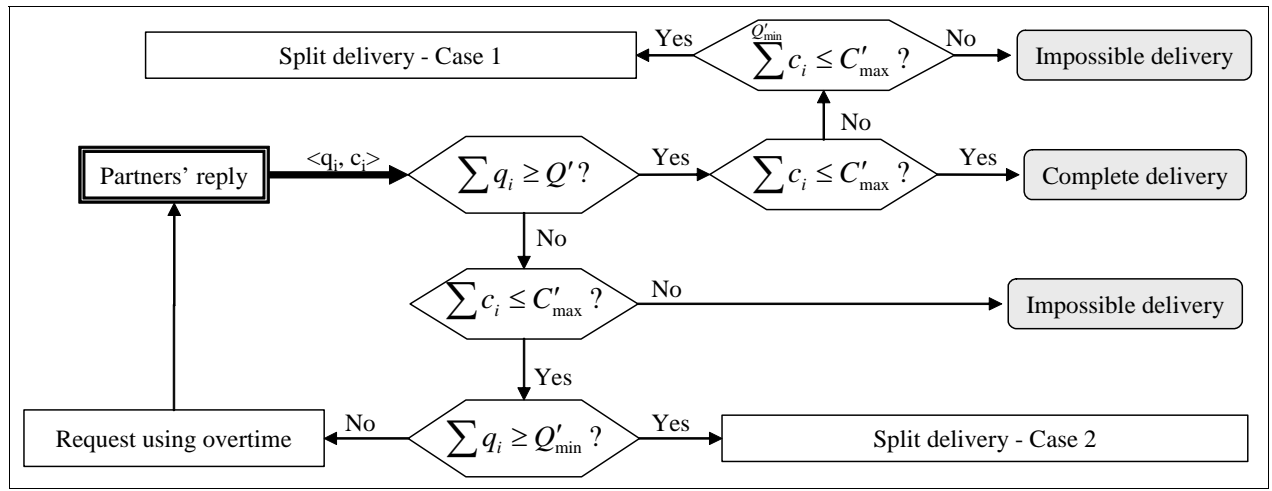

*Figure 5. VSBA decision algorithm*

Then the VSBA has four possible options which are function of partners' answers. It can be the followings:
- Achieve the job by respecting the whole constraints induced by the order.
  And carry out a complete delivery at D.
- Refuse the job, because its induced constraints are too restrictive for the VEN.
- Make a second request allowing overtime.
- Research split delivery possibilities for D$_{max}$ at the latest.

Split delivery consists in supply client the available quantity at D with at least Qmin. (case 2) or Qmin+ (case 1). Qmin+ indicates the sum of the products carried out by the partners retained for the first delivery. Answers not allowing to obtain strictly Qmin, quantity retained for Qmin+ may be higher than Qmin. The missing quantity is delivered for the date Dmax at the latest.

This new request involves an update of the variables of quantity and cost.

$$\text{Case 1: } Q' = Q' - Q'_{min+} \qquad \text{Case 2: } Q' = Q' - \sum q_i \qquad (2)$$

$$\text{Case 1: } C'_{max} = C'_{max} - \sum^{Q'_{min}} c_i \qquad \text{Case 2: } C'_{max} = C'_{max} - \sum c_i \qquad (3)$$

Each member of the VEN could reply to this request by sending its proposal in terms of quantity cost and date (qi, ci, di).

The VSBA analyzes answers of the partners according to the process describes in Figure 6.

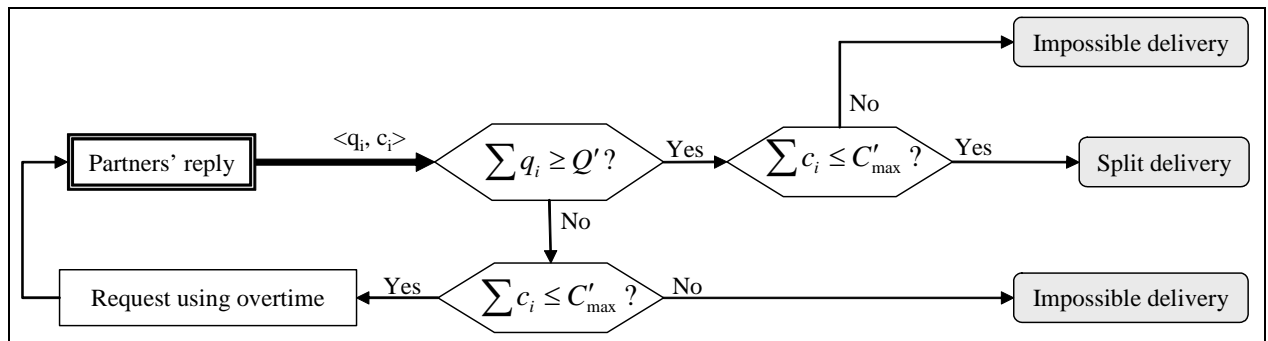

*Figure 6 : Search for a split delivery solution*

### 3.2.3. The single enterprise VEN case

In principle, each VEN is faced with:
- internal constraints related to its capacity limits,
- external constraints, related to its:
  - customer VENs demanding products with a minimal delay at lower cost and at a required quality level,
  - supplier VENs having also constraints of lead times and costs.

The planning and negotiation process of this type of VEN is depicted by Figure 7.

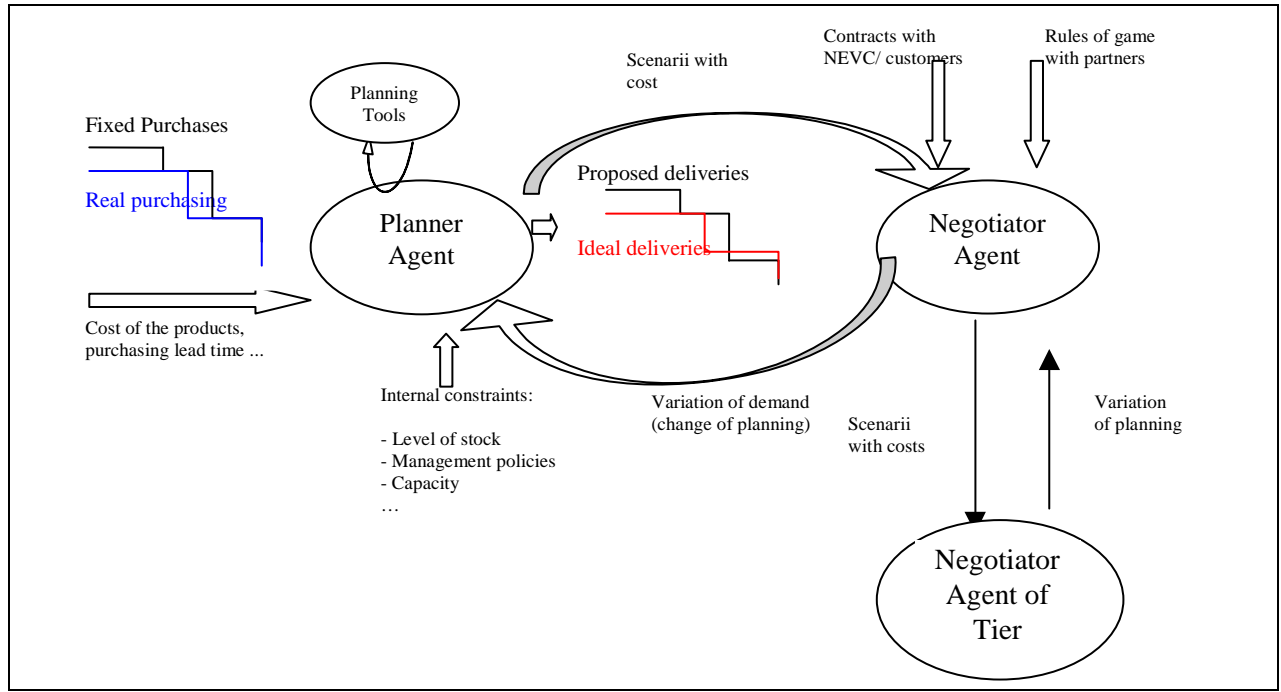

*Figure 7. The planning and negotiation process of a VEN*

To face variations in quantities that the VEN must deliver to its customers, the negotiator agent can reply with the followings:
- I can deliver without any problem
- I can but with addition efforts (using overtime or subcontracting)

- I agree if you fully or partially compensate me with all the over costs, especially if the quantities requested do not respect contracts
- I can supply you - advance future arrivals of components in entry for backward planning - move back the deliveries of certain products for the direct planning.

Otherwise, it is assumed that there is a possibility that the stocks planned for another customer can be delivered to another one which have a great penalties for delays, with compensation of possible over costs.

The NAT takes the set of propositions and chooses propositions with the minimal cost for the associated tier.

### *3.3. Tier level*

#### 3.3.1. Rolling plan and forecasting evolutions

Each VEN of the first tier collects information about future sales from the customers, generally with uncertainty estimation. The forecasts are transmitted to all the VENs of the supply chain. It is assumed on the one hand that agreements are signed between VENs of the first tier and customers, and on the other hand between partners of the SC. Thus, it is assumed that information is always shared truthfully (trusted relationships) (Gavirneni, Kapuscinski, & Tayur, 1999; Cachon, & Lariviere, 2001).

On the basis of forecasting data and contracts established, each VEN makes its planning. In the event of unforeseen production problems of a VEN, or of forecasting change, it must commit with its current customer and supplier VENs, so that they try to overcome the problem collectively, which ensures the continuity of the production and global non stop of the manufacturing chains.

Using a rolling horizon (ouzizi, Anciaux, Portmann, & Vernadat, 2003), the VENs do the planning with new forecasting (Figure 8).

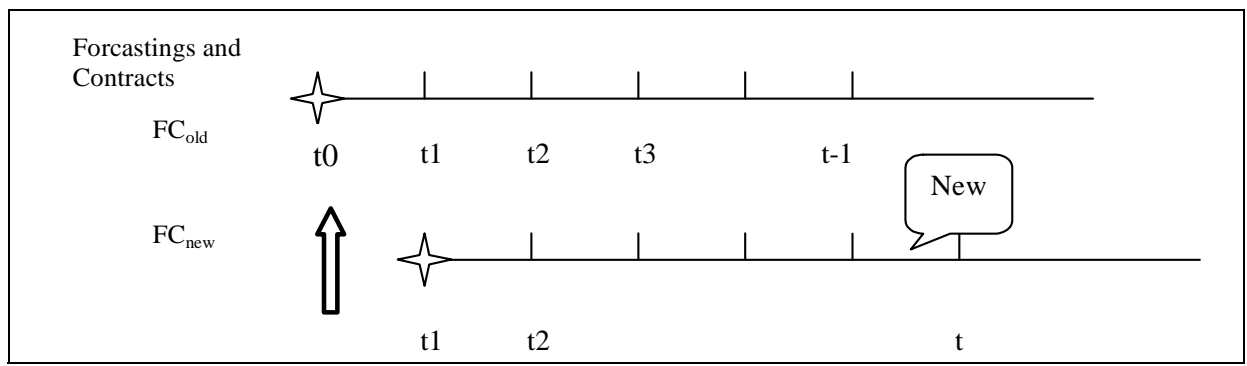

*Figure 8. Rolling horizon for planning*

#### 3.3.2. Planning the supply chain

It is assumed that, (1) in the last unit of time, planning over the supply chain are negotiated and coherent, (2) at the first of the next period of planning, each VEN of the first level must readjust its planning according to the demands variation for finished product, or of possible risks that can occur during the period. Thus, the consequences are:
- to correct forecasts of finished products from the period 1 to T-1 (these forecasts are the same corresponding to periods 2 to T with corrections)
- to add a new forecast for the period T
- to update stocks of all upstream and downstream components of the supply chain.

The problem of each VEN is to determine if it is sufficient to add one period for the planning in order to cope with variations or if it is necessary to change more or less the previous planning so as to find coherent and negotiable planning.

To cope with this problem, the following procedure is proposed:

**A0 -** Different tiers of the supply chain are distinguished, which are denoted by rows. (**NR** is the highest number of rows corresponding to the VENs at the last tier).

**A1 - For each tier from NR down to 1 do**
    *Preliminary work for each tier*
    **For each VEN of the tier do**
    *It is assumed that the due date of products manufactured by the VEN are fixed*
        **For each combination of parameters (costs, penalties, possible delays...) do**
            Construct the backward planning corresponding to the due date deliveries fixed
        **Endfor each combination of parameters**
    **Endfor each VEN**

*For the modification of due date deliveries corresponding to the same component and used by several VENs of one level, the negotiator agent ($NAT_{row}$) of the tier checks if requests anticipated for some nodes are delayed for others. In this case, the problem will not have a serious incidence upstream. Having the sum of all orders and variations, the NAT makes a choice of a planning for each VEN of the tier by taking into account holding costs, production costs corresponding to this tier and perturbations caused.*
*It is thus advisable here to propose a simple method for the choice of planning to be retained (this method can be parameterized to give different results if it is used several times). The definite choice of backward planning makes possible to give the due date deliveries of products manufactured for the adjacent upstream tier.*
**Endfor each l**

**A2 -** In the case where each VEN does not need to place orders of components in the past, the negotiator agent of the tier (NAT) has a set of planning established in A1 that are negotiated and consistent. The NAT selects planning on the basis of the two objectives cited before. If these objectives are satisfied, then everything is ok; otherwise, the A1 procedure starts again with VENs or the NAT changing weights of some penalties.
If backward planning obtained use negative periods, then the VENs establish direct planning. Thus, VENs replace negative times by 0, so that the due dates for purchased products from suppliers of the virtual enterprise are fixed.

**A3 - For tier varying from 1 to NR do**
    *Preliminary work for each tier*
    **For each VEN of the tier do**
        *It is assumed that the due dates of purchased products are fixed*
        **For each combination of parameters chosen by the VEN or the NAT do**
            Construct the direct planning taking into account of products that have been
            manufactured and the values of the two criteria (costs and penalties)
        **Endfor each combination**
    **Endfor each VEN**

*Because it is assumed that any component is provided from one VEN within the SC, it is not possible to obtain compensation between VEN's of the same tier. Thus, in choosing planning, the NAT must not systematically disadvantage always the same VEN and it has to distribute delays over the various customers of each VEN. It is thus advisable here to propose a simple method for the choice of planning (this method can be parameterized to give different results if it is used several times). The definite choice of the direct planning makes possible to give the due date deliveries of components for the adjacent downstream tier.*
**Endfor each row**

**A4** - At the end of step A3, consistent planning of products to be delivered to external customers are obtained but with delays (acceptable or not) as well as the costs and over cost at each VEN. It is then advisable to decide if each one is satisfied by the whole set of consistent planning found and to stop the procedure or if it is necessary to start again in A3 or in A1 by changing penalties or parameters so as to construct a new set of consistent planning.

By limiting the computing time of each direct or backward planning, the number of combinations of parameters and the computing time of the procedures to select planning (number of calls to procedures A1 to A4) are bounded by the number of VENs. Thus, we obtain a global algorithm in which the computing time is bounded by the number of VENs and the number of periods. It is therefore sufficient to call the method for elaborating consistent planning in a limited number of times to find consistent and negotiated planning in a reasonable timeframe.

### 3.4. Supply chain level

The supply chain is modeled as a multi-agent system; agents use cooperative negotiation to establish a coherent planning. To limit the negotiation process in terms of iterations, a negotiator agent for each tier i of the supply chain (NATi) and a mediator agent (MA) for the whole supply chain will be used.

The architecture of the system is given by Figure 9.

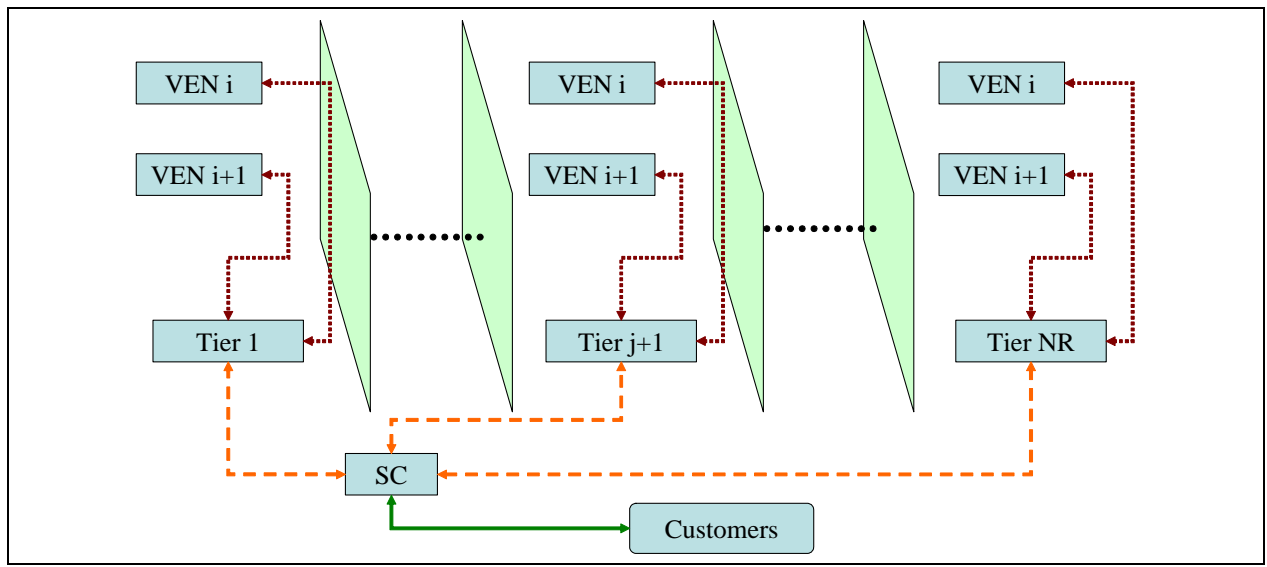

*Figure 9. Global scheme of the virtual enterprise*

## 4. Simple industrial illustration

In this part, we illustrate our approach with the following industrial organization. After organization presentation, an order life cycle is simulated.

### 4.1. Example organization

This example is based on the bronze tap production. In order to facilitate our approach comprehension, we keep only the main components of a sold tap (called PF in Figure 10):
- 1 Bronze body (called SC BBA)
- 2 O ring (called SC BA)
- 1 Blister (called SC A)

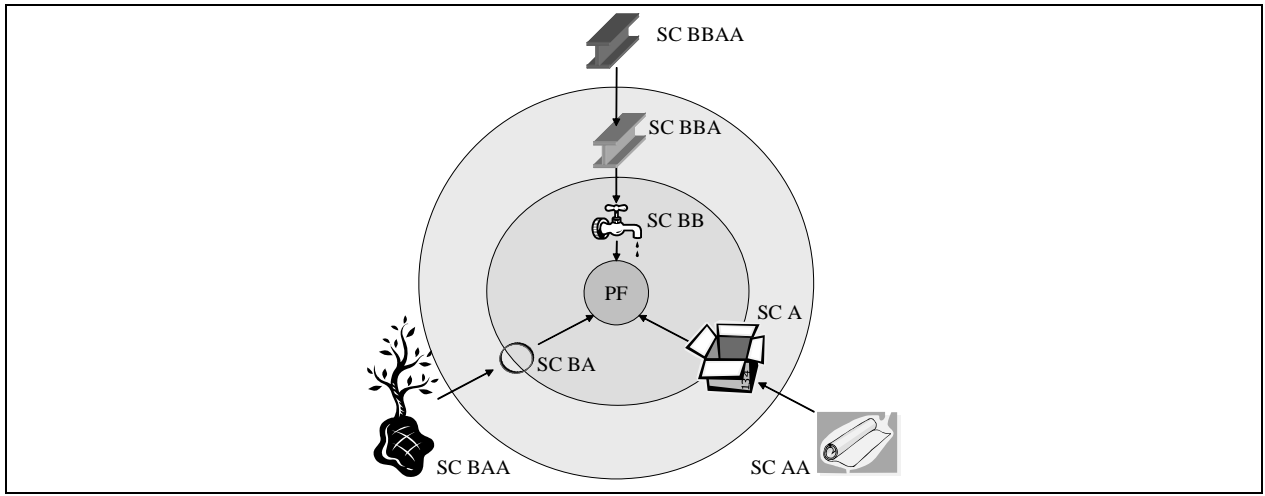

*Figure 10. Sold tap components*

Those components are respectively themselves made of:
- Base metals (copper and tin) (called SC BBAA)
- Rubber (called SC BAA)
- Carton (called SC AA)

All those elements constitute the product breakdown structure (PBS) of bronze tap.

To each element of the PBS corresponds a supplier entity. This entity could be a single enterprise or a regrouping of similar companies. The breakdown structure and corresponding firms allow us to construct the corresponding tap's supply chain. The supply chain is viewed as a set of tiers, in which each partner is in relation with customers and suppliers on adjacent tiers. As Figure 2 and Figure 9 illustrate it, tier relations are supported by the Mediator Agent (MA). All those adjacencies structure supply chain pattern of tiers. This example organization can be summarized as follows (Figure 11).

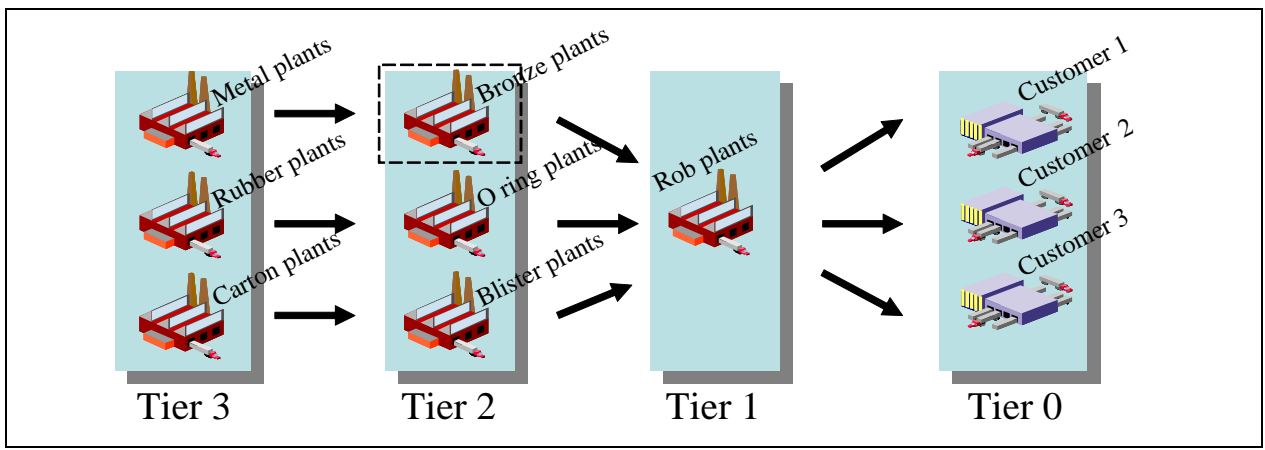

*Figure 11. Supply chain architecture*

We focus our illustration on one VEN constituted by bronze producers (squared on Figure 11). This VEN is, in fact, an association of five bronze plants, in which the Principal Supplier (PS) is the only enterprise visible by network partners.

## 4.2. Order life cycle

The PS receives a new order form *Rob plant* with the following characteristics, which define the Quantities, Delays and maximum Cost allowed by the client: <[1000,750],[50,55],100>. This request provides from the A1 algorithm describes in 3.2.2.

The PS internal analysis concludes than it can only produce 500 unities at a cost of 45 for the given deadline. So, PS sends a message to its VSBA with the following characteristics: <[500,250],[50,55],55>. The VSBA send requests (<500,50,HS=0>) to the four partners to know what their capacities are.

The answers of the partners are:

| Partner | Quantity | Cost |
| --- | --- | --- |
| #1 | 100 | 8 |
| #2 | 60 | 5 |
| #3 | 100 | 12 |
| #4 | 100 | 10 |

In the VSBA, the answers analysis leads to a *Split delivery – Case 2* (cf Figure 5). Indeed, the total quantity is under 500, the total cost is under 55 and the total quantity is upper than 250. For the first delivery, VSBA chooses the following partners, according to their individual performances (Monteiro, & Roy, 2003):

| Partner | Quantity | Cost | Performance | Rank | Retained |
| --- | --- | --- | --- | --- | --- |
| #1 | 100 | 8 | 0,08 | #1 | Yes |
| #2 | 60 | 5 | 0,083 | #2 | Yes |
| #3 | 100 | 12 | 0,12 | #4 | No |
| #4 | 100 | 10 | 0,1 | #3 | Yes |

After those affectations, 240 unities have to be produced at a total cost of 22 for the date 55. So, VSBA sends the request <240,55,HS=0> for the last delivery.

The answers of the partners are:

| Partner | Quantity | Cost |
| --- | --- | --- |
| PS | 30 | 3 |
| #1 | 100 | 8 |
| #2 | 60 | 5 |
| #4 | 40 | 5 |

VSBA use the process described in Figure 6 to analyze those answers. It concludes that over time have to be used. Indeed, total quantity is under 240 and maximum cost is not reached. VSBA sends the request <240,55,HS=1>.

Answers taking into account overtime are:

| Partner | Quantity | Cost |
|---|---|---|
| PS | 70 | 8 |
| #1 | 150 | 13 |
| #2 | 90 | 8 |
| #3 | 40 | 6 |
| #4 | 90 | 12 |

VSBA analysis concludes to a possible split delivery. Indeed total quantity is could reach 240 without exceed the maximum cost of 22. So, for the last delivery, VSBA chooses the following partners:

| Partner | Quantity | Cost | Performance | Rank | Retained |
|---|---|---|---|---|---|
| PS | 70 | 8 | 0,114 | #4 | No |
| #1 | 150 | 13 | 0,083 | #1 | Yes |
| #2 | 90 | 8 | 0,086 | #2 | Yes |
| #3 | 40 | 6 | 0,15 | #5 | No |
| #4 | 90 | 12 | 0,13 | #3 | No |

Finally, this order could be supported by the VEN with the following work distribution:

| Partner | First Delivery (Date 50) | | Last Delivery (Date 55) | |
|---|---|---|---|---|
| | Quantity | Cost | Quantity | Cost |
| PS | 500 | 45 | 0 | 0 |
| #1 | 100 | 8 | 150 | 13 |
| #2 | 60 | 5 | 90 | 8 |
| #4 | 100 | 10 | 0 | 0 |
| **Totals** | **760** | *68* | **240** | *21* |

With those two deliveries, the total quantity is 1000 unities at a cost of 99.

This work distribution is send to Tier 2 negotiator agent to analyze coordination problem using also *O ring plant* and *Blister plant* responses (A1 algorithm). If Tier 2 does not generate trouble, all those information are forwarded to Tier 1 partner: *Rob plant*. On the contrary, if coordination problem is detected, Tier 2 negotiator agent (using A3 algorithm) send modification request to upper Tier (*Rob plant*).

## 5. Conclusion and future research

New industrial architecture organization, based on cooperation, highlight the problem of flow control and management by independent decision centers. Decision distribution along the supply chain needs coherence between partners in order to achieve better productivity and greater reactivity.

The elementary actors of the supply chain that we had described here could have two different forms. First, it could be a single enterprise in which a Planner Agent allows a dynamical planning based on the global cost of the supply chain. So each partner is able to elaborate planning according to unforeseen orders, assuming that quality-cost-delay framework contracts have previously been drawn up between the various partners of the supply chain. Second, the virtual enterprise could be constituted by many autonomous enterprises with similar productions. This second approach involves one agent by enterprise and a Virtual Space Broker Agent helping to the load distribution. On the supply chain view, those two approaches are analogous.

Theses actors integrate an architecture guarantying, in the same time, robustness and agility of the supply chain.

The Negotiator Agent, in relation with the tier level, and Mediator Agent, in relation with the supply chain level, permit the use of rolling plan and limit the negotiation process in terms of iterations.

Some evolutions of this system which only concerns a supply network, could be extend to the distribution network. Moreover, the multi-sourcing (multi-VEN) possibilities are not implemented yet and could be studied further.

## 6. Acknowledgements